# Magnetostriction and magnetic texture to 97.4 Tesla in frustrated SrCu$_2$(BO$_3$)$_2$

Marcelo Jaime [a,b,†], Ramzy Daou [c], Scott A. Crooker [a,b], Franziska Weickert [b], Atsuko Uchida [a,b], Adrian Feiguin [d], Cristian D. Batista [e], Hanna A. Dabkowska [f] and Bruce D. Gaulin [f,g].

[a] NHMFL, Los Alamos National Laboratory, Los Alamos, New Mexico 87544, USA
[b] MPA-CMMS, Los Alamos National Laboratory, Los Alamos, New Mexico 87544, USA.
[c] Max Planck Institute for Chemical Physics of Solids, 01187 Dresden, Germany
[d] Department of Physics & Astronomy, University of Wyoming, Laramie, Wyoming 82071, USA
[e] Theory Division, Los Alamos National Laboratory, Los Alamos, New Mexico 87544, USA
[f] Brockhouse Institute for Materials Research, McMaster University, Hamilton, ON, L8S 4M1, Canada
[g] Department of Physics & Astronomy, McMaster University, Hamilton, ON, L8S 4M1, Canada

**Strong geometrical frustration in magnets leads to exotic states, such as spin liquids, spin supersolids and complex magnetic textures. SrCu$_2$(BO$_3$)$_2$, a spin-1/2 Heisenberg antiferromagnet in the archetypical Shastry-Sutherland lattice, exhibits a rich spectrum of magnetization plateaus and stripe-like magnetic textures in applied fields. The structure of these plateaus is still highly controversial due to the intrinsic complexity associated with frustration and competing length scales. We reveal new magnetic textures in SrCu$_2$(BO$_3$)$_2$ via magnetostriction and magnetocaloric measurements in fields up to 97.4 Tesla. In addition to observing the low-field fine structure of the plateaus with unprecedented resolution, the data also reveal lattice responses at 82 T and at 73.6 T which we attribute, using a controlled density matrix renormalization group approach, to the long-predicted 1/2-saturation plateau, and to a new 2/5 plateau.**

Quantum paramagnets are Mott insulators where dominant intra-cell antiferromagnetic interactions lead to a singlet ground state that remains stable for small inter-cell interactions. The low energy spectrum of triplet excitations is gapped. For systems with uniaxial symmetry, the gap ($\Delta$) is closed by applying a magnetic field $g\mu_0H_0 \approx \Delta$ along the symmetry z-axis (see Figs.

1A and 1B). The low energy degrees of freedom can be mapped onto a gas of hard-core bosons (1). In this description, the external magnetic field plays the role of a chemical potential, and the longitudinal magnetization corresponds to the boson concentration. In systems such as BaCuSi$_4$O$_6$ (2), where Cu atoms are arranged in parallel dimers on a square lattice, the triplons condense in a phase-coherent fluid analogous to a Bose-Einstein condensate. On the other hand, if the kinetic energy is highly frustrated, the repulsion between triplets becomes dominant and leads to crystals with superstructures that are highly sensitive to the concentration of triplets or magnetization (3). These crystalline states correspond to Ising-like orderings, *i.e.*, states with spontaneous modulation of the spin component parallel to the field. The orthogonal-dimer geometry of the Shastry-Sutherland lattice (SSL) (Fig. 1A) (4) and the ratio of next-nearest to nearest neighbor exchange interactions between the spin 1/2 Cu$^{2+}$ ions, $J_1/J_0 \sim 0.62$ ($J_0 \approx 74K$), make SrCu$_2$(BO$_3$)$_2$ a paradigm of frustrated quantum magnetism (5-8). Nuclear magnetic resonance (9) and torque magnetometry (10,11) measurements have been used in the past to study the magnetic superstructures in SrCu$_2$(BO$_3$)$_2$, but the strength of required magnetic fields has prevented the unambiguous observation of magnetization fractions beyond 1/3 of saturation. Sensitive measurements for higher concentrations are crucial for model validation. Indeed, given the limited lattice sizes for which the minimal model for SrCu$_2$(BO$_3$)$_2$ can be solved under control, it is very important to have experimental data for magnetic fields that minimize the unit cell size of the magnetic superstructures. This condition is satisfied by the 1/2 plateau that becomes stable at extremely high fields.

Given the difficulties associated with conventional magnetization measurements in pulsed magnetic fields, spin-lattice coupling provides an alternative and attractive route to probe magnetization and magnetic interactions in quantum magnets via their influence on the sample size. To this end, Daou et al. (12) recently developed a very sensitive, all-optical technique based on fiber Bragg gratings to measure magnetostriction (MS) in ultrahigh pulsed magnetic fields, thereby enabling strain measurements of materials that require high magnetic fields to overcome strong antiferromagnetic spin interactions. Here we combine five essential advances, used to reveal new magnetic textures in $SrCu_2(BO_3)_2$. Namely, we *(i)* measure simultaneously the magnetostriction and magnetocaloric effect, with *(ii)* the sample exactly aligned with the applied magnetic field (in contrast with the limitations of torque magnetometry to only magnetically anisotropic materials). We also extend the measurement range to *(iii)* sub-Kelvin temperatures (where transitions are sharper), in *(iv)* a record high nondestructive magnetic fields up to 97.4 T (required to achieve 1/2 saturation). Finally, *(v)* we expand the use of a DMRG numerical technique to 4xL sites to study magnetic textures beyond 1/3 in $SrCu_2(BO_3)_2$.

Fig. 1C shows the linear magnetostriction ($\Delta L/L$) of $SrCu_2(BO_3)_2$ measured at 0.58 K using a 50T pulsed magnet ($\Delta L//H//c$). For comparison, the directly-measured magnetization is also shown in Fig. 1D (from Ref. 8). The curves demonstrate a remarkable correspondence between magnetostriction and spin ordering. Both evolve similarly with decreasing temperatures (13), both increase smoothly when the spin gap closes at ≈20T at low temperatures, and both increase sharply at the onset of the primary 1/9, 1/4, and 1/3 magnetization plateaus at 26.9, 36.0, and 39.8 T, respectively. Moreover, considerable fine structure is reproducibly observed between the 1/9 and 1/4 plateaus (see inset), where the 1/8, 1/7, 1/6, 1/5, and even a hint of the 2/9 plateau are revealed, confirming both theoretical prediction (10,14,15) and prior magnetometry (10,11), and demonstrating the micro-strain sensitivity of the MS technique.

Fig. 1E shows the simultaneously-measured magnetocaloric effect (MCE); that is, the field-dependent changes in sample temperature due to changes in magnetic entropy or to dissipation. Pronounced MCE exists when the spin gap closes at 20T, similar to results in the magnon BEC system $Ba_3Mn_2O_8$ (16). A strong irreversible MCE also exists at the 1/3 plateau, consistent with a first-order magnetic phase transition. The absence of measurable MCE between the 1/9 and 1/4 plateaus might be due to the more gradual magnetization increase through closely spaced superstructures that are proposed to coexist with an itinerant triplet component that propagates along the interstitial sites of the crystal (13,17).

To explore the existence and stability of additional magnetic textures in $SrCu_2(BO_3)_2$, we performed MS in ultrahigh magnetic fields to 97.4 T. The data (Fig. 2) reveal two new features beyond the 1/3 magnetization plateau, thereby providing the first unambiguous evidence for new magnetic structures at high fields: an increase in $\Delta L$ at 73.6 T, followed by a sharp drop at 82.0 T. The latter is assigned to the onset of a stable magnetization plateau at 1/2 saturation, long predicted by theory and first hinted at in recent magnetization studies (10). The former feature is entirely new, and numerical studies described below strongly suggest the formation of a stable magnetic texture at 2/5 saturation.

Numerical calculations of stable magnetic textures become more controlled for high triplet concentrations. Most calculations suggest simple stripe structures for the 1/3 and 1/2 plateaus (18). The insets of Fig. 3A depict the $M/M_{sat}$ = 1/2 structure as alternating triplet and singlet columns (*ts*), while the $M/M_{sat}$= 1/3 ordering has two singlet columns separating triplet columns (*tss*). Assuming that the lowest-energy textures have one-dimensional stripe patterns in this field range, and assuming exponentially-decaying repulsive stripe interactions (18), a stable 2/5 plateau with alternating 1/3 and 1/2 stripe pattern – *tstss* (see inset) -- is expected. At intermediate fields, this texture is favored because it avoids triplet columns separated by 4 and 6 lattice spaces. To be more precise, if $V_j$ is the effective repulsion between triplet stripes separated by a distance *j* (see Fig.3A), the field interval $\Delta H$ over which the 2/5 plateau is stable is given by $g_c\mu_B\Delta H=5[(V_4+V_6)-(V_5+V_7)]$ (neglecting $V_j$ for $j \geq 8$). It is clear that the 2/5 plateau is stable ($\Delta H>0$) because $V_4>V_5$ and $V_6>V_7$. In simple terms, triplet stripes are separated by 5 and 7 lattice spaces in the 2/5 plateau, so the corresponding energy cost is lower than the energy gain. This simple argument is confirmed by a DMRG study of the Shastry-Sutherland Hamiltonian in clusters of $L_y$ x $L_x$ spins with $L_y$ = 4, $L_x$ = 30, $J_0$ =78K and $J_1$ = 0.62 $J_0$. The calculated $M(H)/M_{sat}$ curves (Fig. 3A) show very stable 1/3, 2/5, and 1/2 plateaus occurring at high fields. Fig. 3b displays the two-point correlation function of the spin component parallel to H versus the distance along the *x*-axis, confirming the pattern expected for the 2/5 plateau. We note the excellent agreement between the calculated and measured critical fields for the 1/3, 2/5 and 1/2 plateaus. In addition, we cannot exclude finer structures between the 1/3, 2/5 and 1/2 plateaus that could appear for bigger cluster sizes.

Finally, we infer the volumetric MS ($\Delta V/V$), which nominally requires knowledge of $\Delta L/L$ along all crystal axes (*a*, *b*, and *c*). We therefore measured MS in the *ab*-plane ($\Delta L//H$, $H\perp c$). Fig. 4 shows $\Delta L/L$ measured in a 60T pulsed magnet at 1.5K at the HLD (Dresden), and also in a 65T pulsed magnet at 0.5K at the NHMFL (Los Alamos) for both $H//c$ and $H\perp c$. We then scale the field axis by the anisotropic *g*-factor ($g_{//c}$ = 2.28 and $g_{\perp c}$ = 2.04) (19). See MS measured for $H \perp c$ at other temperatures in the Supplementary Figure 3. While the *H//c* data are very consistent across experimental platforms, the $H \perp c$ data differ at the 1/3 plateau, where broken tetragonal symmetry can play a role (*a*- and *b*- axes become inequivalent upon stripe formation). Our *ΔL//H⊥c* data agree in both magnitude and sign with X-ray data taken for *H//c*, *ΔL//a* (13,20); therefore we temporarily assume only *g*-factor anisotropy and compute *ΔV/V* (Fig. 4). We assume (arbitrarily) that Δ*a*/*a* is measured at HLD and that Δ*b*/*b* is measured at the NHMFL. Figure 4 shows that at low fields <45 T, *ΔV/V* is largely independent of the calculation method ($a^2c$, *abc*, or $b^2c$,), and the crystal volume shrinks with *H* because oxygen-mediated Cu-O-Cu superexchange favors an angle closer to 90° when the Cu spins are parallel causing the reduction of the intradimer Cu-Cu bond (20). However, this trend stops once the density of triplets exceeds 1/3.

In summary, we studied magnetization and spin-lattice correlations in $SrCu_2(BO_3)_2$ using microstrain-resolved magnetostriction measurements in ultrahigh pulsed magnetic fields to 97.4 T. Two sharp transitions at 73.6 and 82 T are attributed to a new stable 2/5 magnetization plateau, and to the long-predicted 1/2 plateau respectively. Numerical calculations indicate very stable 1/3, 2/5, and 1/2 magnetization textures, in agreement with the data. Estimates of the unit cell volume indicate shrinkage up to the 1/4 plateau, after which expansion occurs, likely due to strong triplet repulsion. Millikelvin-resolved magneto-caloric effect measurements show evidence for dissipation and and irreversibility at the 1/3 plateau, as expected for first-order phase transitions. These magnetostriction studies therefore provide a reliable route toward high-sensitivity measurements of magnetic order in the very high magnetic fields necessary to overcome strong magnetic interactions, paving the way for elucidating textures in the ever-more interesting family of magnetically frustrated quantum magnets.

Materials and Methods

Single crystal growth: High quality single crystals of incongruently melting $SrCu_2(BO_3)_2$ were grown by optical floating zone image furnace techniques using self-flux. Single crystals were oriented and characterized by X-ray diffraction and by neutron scattering using highly enriched $^{11}B$ for selected crystals as described by Dabkowska et al. (21).

Magnetostriction: A single-mode optical fiber containing a 1 mm long fiber Bragg grating (FBG) manufactured by SmartFibre, UK, is attached to the sample using cyanoacrylate bond. Broadband light (1525–1565 nm) from a superluminescent diode illuminates the FBG. When the sample expands or contracts, the narrow band of light (≈1550 nm) that is reflected from the FBG shifts slightly. This reflected light is dispersed in a 500 mm spectrometer with a 600 groove/mm diffraction grating, and its spectrum is detected by a fast 1024-pixel InGaAs line array camera that is read out at 47 kHz. The setup is adapted from that originally demonstrated by Daou et al. (12).

Magnetocaloric effect: To measure the sample temperature as a continuous function of magnetic field, a Cernox® CX1010 bare chip thermometer is bonded directly to the sample using GE-7031 varnish. The resistance of the thermometer is monitored at 100 kHz using a home-built digital lock-in amplifier (22). In the pulsed field experiments the sample is immersed in liquid $^4He$ or liquid $^3He$, and a good but finite thermal link to the bath is provided by the thermal conductivity of the sample itself.

DMRG: We have used the density matrix renormalization group method (23-25) (DMRG) to simulate the Shastry-Sutherland model on a 4xL lattice. We have chosen the length L=30 to be commensurate with the expected ordered states at magnetizations $M/M_{sat}$=1/3, 2/5, 1/2, and periodic boundary conditions to avoid boundary effects that can be very strong in these ordered phases. Notice that the Hilbert space for this system has the same dimension as a 2xL Hubbard ladder. Even though this choice of boundary conditions makes the simulation more lengthy, we have found that in practice, due to the weakly entangled nature of the ground states, we need to keep only about 1200 DMRG basis states to achieve good convergence in the energy and correlations. In order to optimize convergence (26), we have followed a one-dimensional path passing through the strong (diagonal) bonds. To pin the order and avoid the effects of irrelevant degeneracies, we have added an external magnetic field at the boundary for the first four sweeps.

Acknowledgements

We thank C. Swenson, D. Rickel, D. Roybal, M. Gordon, J. Martin and Y. Coulter for the operation of the NHMFL 100 T magnet, J. Betts and F. Balakirev for assistance with experiments, and to N. Harrison for the determination of the magnetic field using de Haas-van Alphen effect in copper. MJ thanks F. Steglich and M. Nicklas for their hospitality at the MPI-CPfS, Dresden. AF acknowledges NSF funding under grant DMRG-0955707. Experiments at the High Magnetic Field Laboratory Dresden (HLD) were sponsored by Euro-MagNET II under the EU contract 228043. Work at the NHMFL was supported by the National Science Foundation, the US Department of Energy, and the State of Florida.

Author contributions: M.J., R.D. Designed Research, M.J., R.D., S.A.C., F.W., and A.U. performed research. M.J., R.D., S.A.C., F.W., C.D.B. and A.F. analyzed data, H. D. and B.G. grew samples. M.J., R.D., S.A.C., F.W., C.D.B. and A.F. wrote the manuscript.

The authors declare no conflict of interests.

This article is a PNAS Direct submission.

[†]To whom correspondence should be addressed: mjaime@lanl.govReferences

1. Giamarchi T, Tsvelik (1999) Coupled ladders in a magnetic field. *Phys. Rev. B* 59:11398-11407.

2. Jaime M, et al. (2004) Magnetic field induced condensation of triplons in Han Purple pigment $BaCuSi_2O_6$. *Phys. Rev. Lett.* 93:087203/1-4.

3. Rice TM (2002) To condense or not to condense. *Science* 298:760-761.

4. Shastry BS, Sutherland B (1981) Exact ground-state of a quantum-mechanical antiferromagnet. *Physica B+C* 108:1069-1070.

5. Kageyama H, et al. (1999) Exact dimer ground state and quantized magnetization plateaus in the two-dimensional spin system $SrCu_2(BO_3)_2$. *Phys. Rev. Lett.* 82:3168-3171.

6. S. Miyahara S, Becca F, Mila F (2003) Theory of spin-density profile and lattice distortion in the magnetization plateaus of $SrCu_2(BO_3)_2$. *Phys. Rev. B* 68:024401/1-10.

7. Gaulin BD, et al. (2004) High-resolution study of spin excitations in the singlet ground state of $SrCu_2(BO_3)_2$. *Phys. Rev. Lett.* 93:267202/1-4.

8. Jorge GA, et al. (2005) Crystal symmetry and high-magnetic-field specific heat of $SrCu_2(BO_3)_2$. *Phys Rev. B* 71:092403/1-4.

9. Kodama K, et al. (2002) Magnetic superstructure in the two-dimensional quantum antiferromagnet SrCu2(BO3)2. *Science* 298:395-399.

10. Sebastian SE, et al. (2008) Fractalization drives crystalline states in a frustrated spin system. *Proc. Nat.l Acad. Sci.* 105:20157-20160.

11. Levy F, et al. (2008) Field dependence of the quantum ground state in the Shastry–Sutherland system $SrCu_2(BO_3)_2$. *Europhys. Lett.* 81:67004/p1-p5.

Figure Labels

**Fig. 1. (A)**, A depiction of the spin-1/2 $Cu^{2+}$ atoms in the Shastry-Sutherland lattice, as realized in $SrCu_2(BO_3)_2$. $J_0$ and $J_1$ are the competing antiferromagnetic exchange constants ($J_0$=74 K, $J_1/J_0$ = 0.62) causing magnetic frustration. **(B)** Singlet and triplet levels of the spin dimers are separated by an energy gap $\Delta$ in zero magnetic field. A magnetic field $g\mu_0H_0 \approx \Delta$ can close this spin gap, Zeeman effect, inducing the formation of triplets and magnetic order. Also shown is a non-zero Dzyaloshinski-Moriya (DM) term due to buckling of the $Cu^{2+}$ planes, which can cause a triplet level splitting at $H = 0$ and anti-crossing at $\mu_0H_0$. **(C)** Magnetostriction ($\Delta L/L$) versus magnetic field measured for $\Delta L$ // $H$ // $c$-axis during the upsweep (red) and downsweep (blue) of a 45 T pulsed magnetic field. **(D)** The magnetization vs field shows the closing of the spin gap at ≈20T, and plateaus corresponding to 1/9, 1/4, and 1/3 of magnetization saturation. Inset: $\Delta L/L$ expanded in the 28-34 T range, which also reveals features corresponding to 1/9, 1/8, 1/7, 1/6, 1/5, and 2/9 of saturation magnetization. **(E)** The magnetocaloric effect (MCE), measured simultaneously with magnetostriction, for field upsweep (orange) and downsweep (green).

**Fig. 2.** Magnetostriction vs magnetic field for $\Delta L$ // $H$ // $c$-axis measured in 50T (red) and 100T (blue) pulsed magnets. Data taken during field upsweep and downsweep are included. We observe a sharp increase in the sample length at 39.7T, corresponding to the onset of the 1/3

magnetization plateau in both data sets. Two new features at 73.6 T and 82.0 T in the high field data are attributed to the onset of the 2/5 and 1/2 magnetization plateaus. Two datasets (shifted for clarity) demonstrate good reproducibility. Inset: field profile of the 100T repetitively-pulsed magnet at the NHMFL.

Fig. 3. (A), Normalized magnetization vs field calculated in a 4x30 site lattice for $J_0$ = 78K and $J_1/J_0$ = 0.62 showing that the most stable magnetization plateaus occur at 1/3, 2/5, and 1/2 of magnetization saturation. (B) Spin correlation function calculated for the 2/5 plateau showing a stable superstructure.

Fig. 4. (A) Magnetostriction vs scaled magnetic field for $\Delta L$ // $H$ // c-axis (blue, left axis), and $\Delta L$ // $H \perp$ c-axis (green curves, right axis) measured in pulsed magnets at NHMFL (Los Alamos) and HLD (Dresden). Values used for the anisotropic g-factor are $g_{//c}$ = 2.28 and $g_{\perp c}$ = 2.04. The green curves show shrinking of the lattice in the ab plane (B) Volumetric magnetostriction $\Delta V/V$ calculated assuming only g-factor anisotropy. The red curve was computed additionally assuming that $\Delta L$ // $H \perp$ c-axis represents the behavior of the crystal for $H$ // b-axis and labeled (axbxc). Black lines were computed assuming that either $\Delta L$ // $H \perp$ c-axis data can represent the a- or b-axis response in the high field orthorombic structure and labeled ($a^2$xc) and ($b^2$xc) respectively. Dashed lines are guides to the eye.

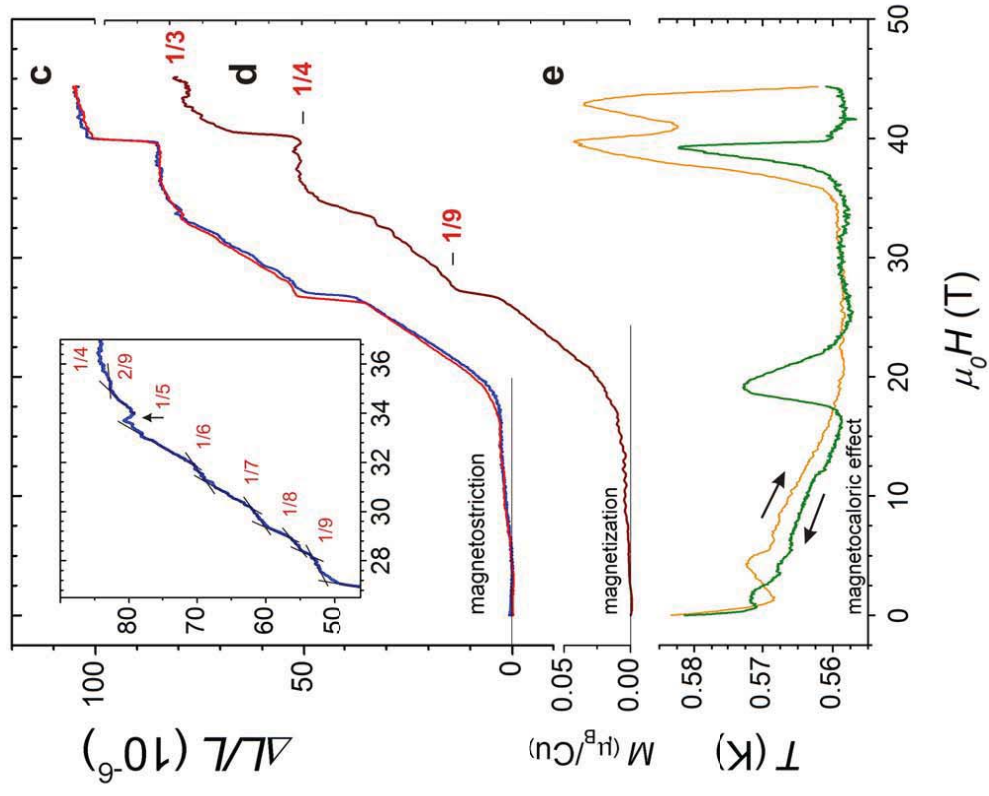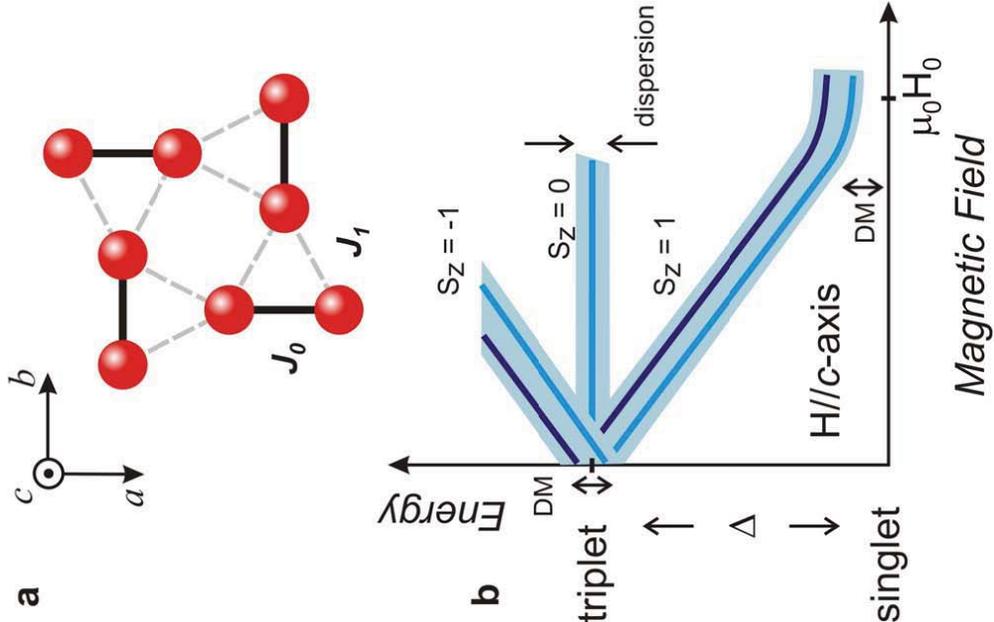

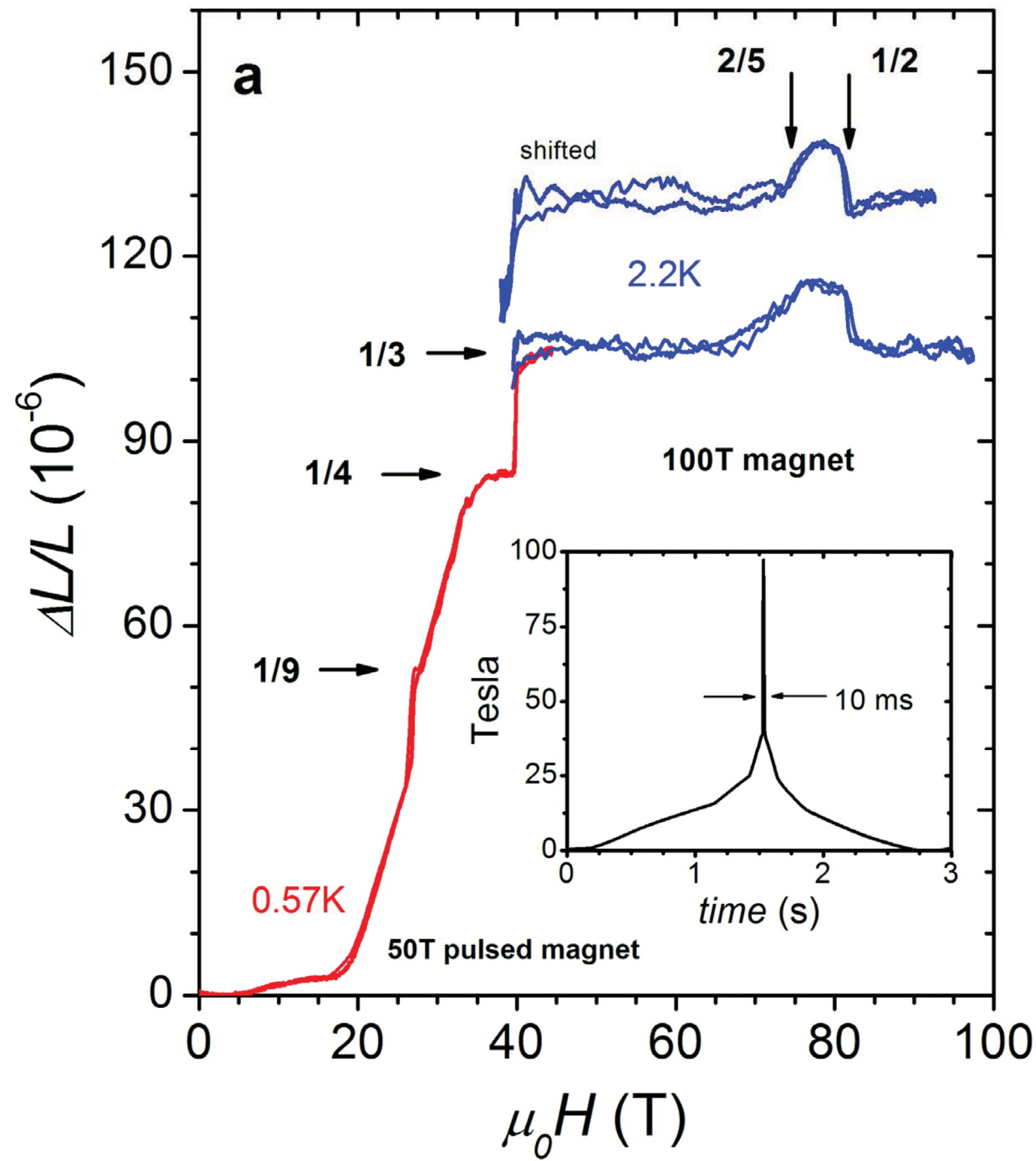

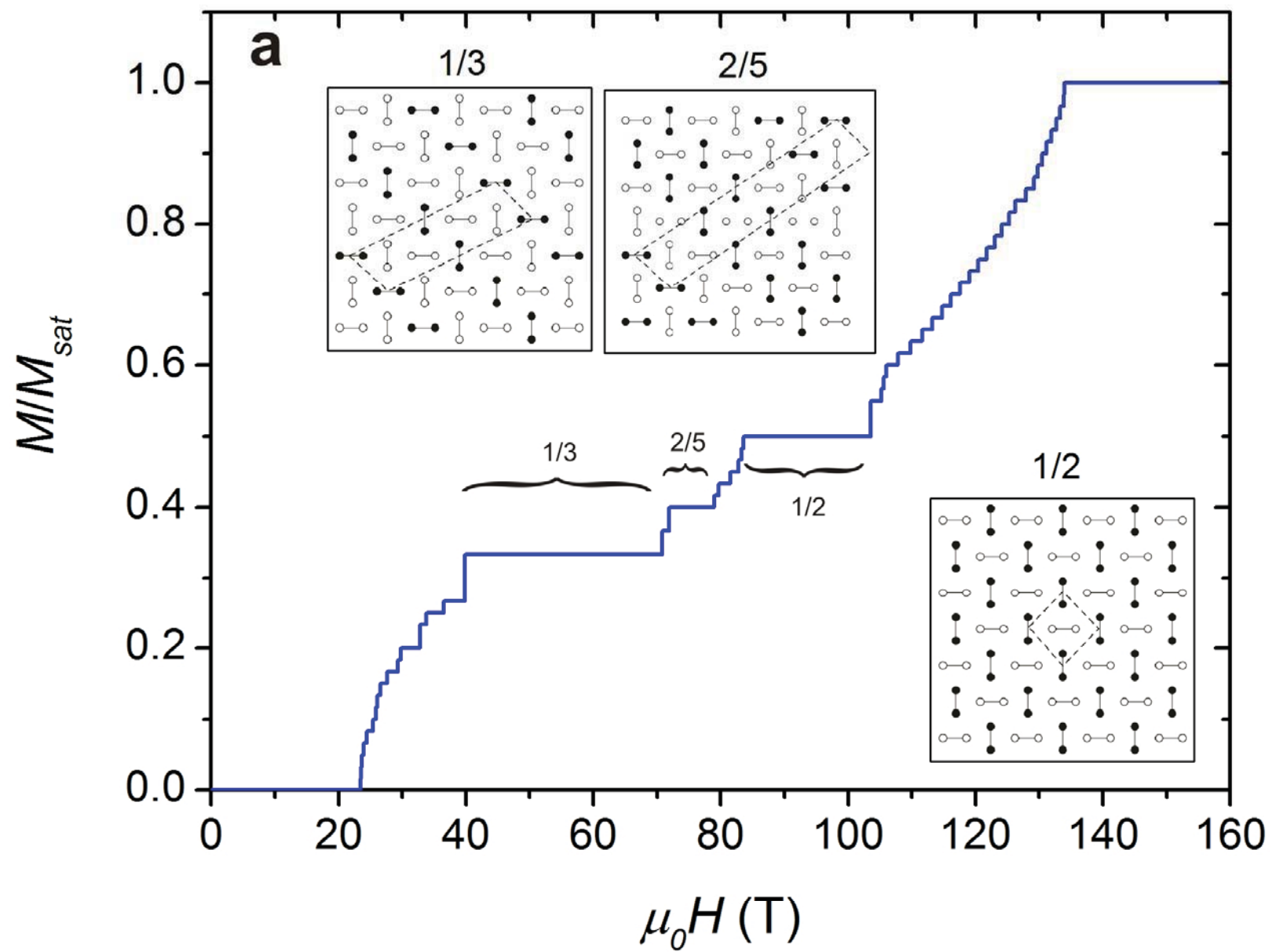

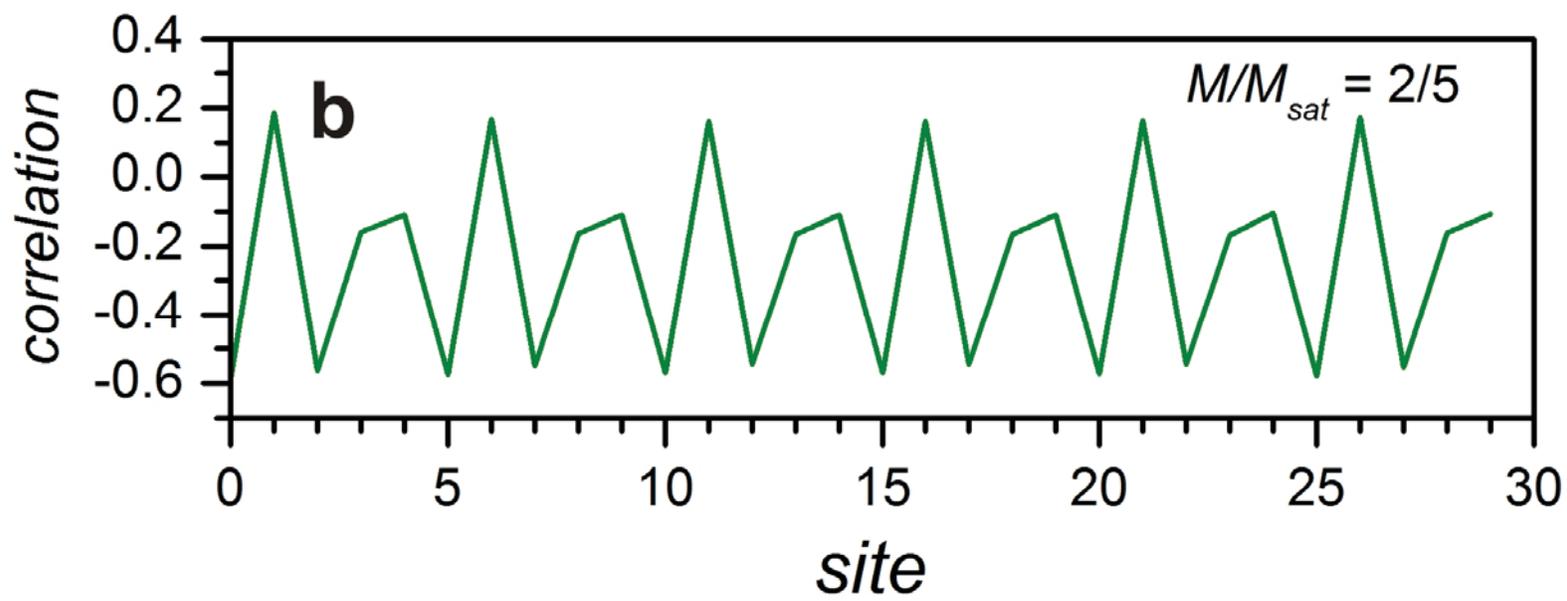

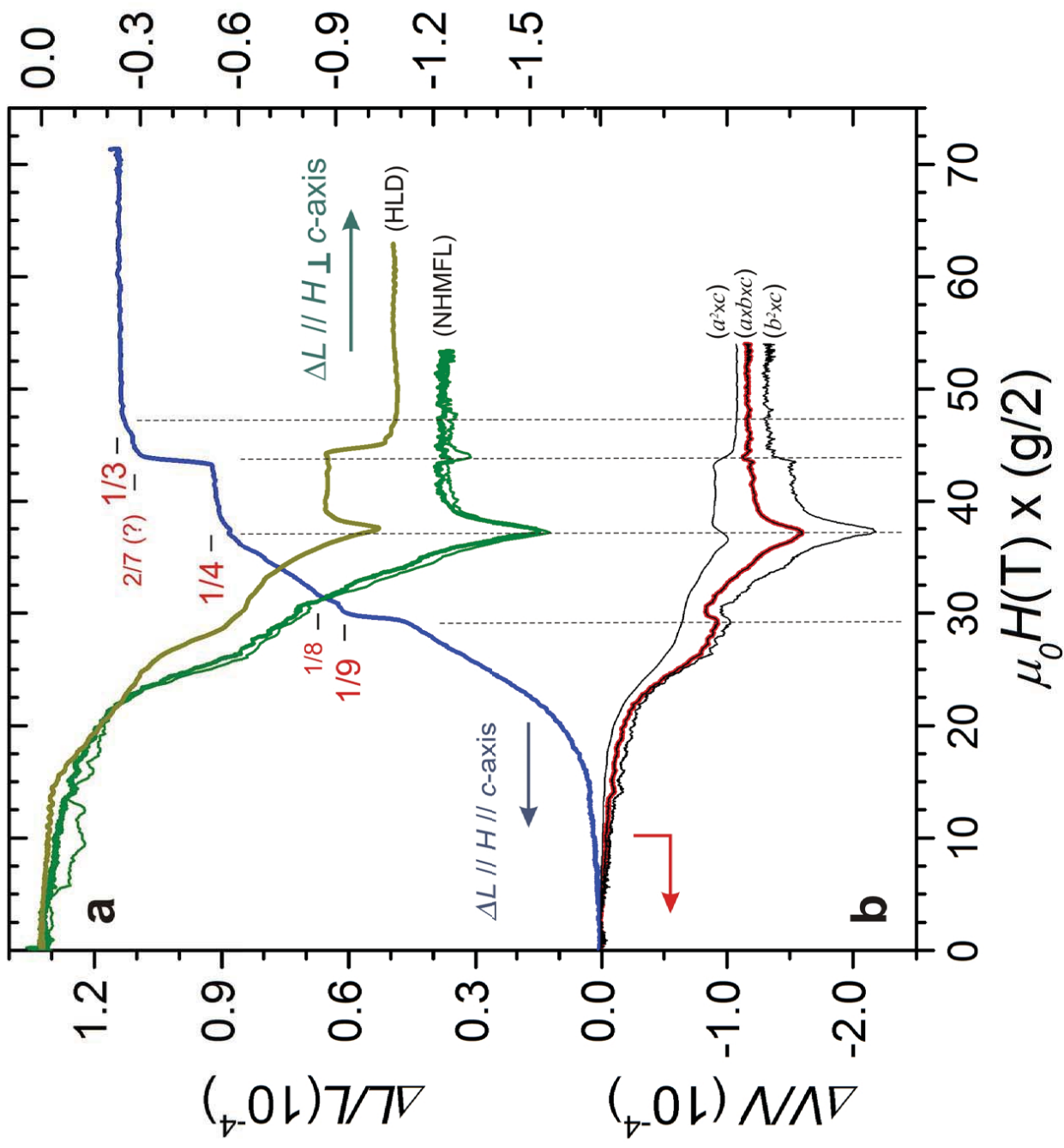